# Generation of above-TW 1.5-cycle visible pulses at 1 kHz by post-compression in a hollow fiber

Tamas Nagy,[1,*] Martin Kretschmar,[1] Marc J. J. Vrakking,[1] Arnaud Rouzée[1]

[1]*Max Born Institute for Nonlinear Optics and Short Pulse Spectroscopy, Max-Born-Straße 2A, 12489 Berlin, Germany*
*Corresponding author: nagy@mbi-berlin.de

**We report on the generation of 6.1 mJ, 3.8 fs pulses by the compression of a kHz Ti:sapphire laser in a large-aperture long hollow fiber. In order to find optimal conditions for spectral broadening at high pulse energies, we explore different parameter ranges where ionization or the Kerr effect dominates. After identifying the optimum parameter settings, large spectral broadening at high waveguide transmission is obtained. The intense 1.5-cycle pulses are used for high-harmonic generation in argon and neon. © 2020 Optical Society of America**

http://dx.doi.org/10.1364/OL.99.099999

More than two decades ago first spectral broadening experiments in gas-filled hollow-core fibers (HCFs) [1] quickly led to the generation of few-cycle pulses at sub-mJ-levels [2], marking more than three orders of magnitude increase in pulse energy compared to the former state-of-the-art utilizing single-mode silica fibers [3]. This dramatic energy boost made these pulses suitable for strong-field applications, such as isolated attosecond pulse generation. Since the pioneering days there is a continuous quest towards further increase of the pulse energy and peak power of few-cycle pulses, driven by demanding applications such as attosecond nonlinear spectroscopy [4] and electron acceleration [5].

However, the energy scaling of HCF compressors has proven to be challenging due to several physical reasons [6]: (i) In order to maintain ideal conditions for Kerr-effect-induced spectral broadening, self-focusing in the waveguide needs to be prevented. Therefore, as the input peak power is increased, the nonlinearity of the medium needs to be decreased. This can be solved by using light gases such as helium or neon and applying circularly polarized light, which further reduces the strength of the nonlinear interaction by about 33% [7]. (ii) In order to minimize losses it is also desirable to prevent ionization, which limits the applicable peak intensity. Large waveguide diameters and long HCF allows maintaining large spectral broadening while keeping the peak intensity moderate [8]. However, such long waveguide geometries have only recently come available with the introduction of stretched flexible hollow fibers (SF-HCF) [9]. Additionally, an increase of the input peak power without dramatic degradation of the transmission due to filamentation in front of the fiber can be achieved by evacuating the front side of the HCF and using a pressure gradient along the waveguide [10].

An alternative approach to Kerr-induced spectral broadening is the exploitation of ionization-induced phase modulation [11]. This approach allows using significantly higher input pulse energies compared to Kerr-based techniques, at the expense of much higher transmission losses making it most suitable for 100-mJ-class lasers. In this way compressed pulses with more than 10 mJ have been demonstrated [12], however the pulse duration could not be pushed below ~9 fs [13] limiting their applications.

Filamentation was also considered as a potential tool for high-energy few-cycle pulse generation [14] where self-compression dynamics could also be demonstrated [15-17]. However, the complex coupling of several nonlinear processes makes a well-controlled up-scaling very challenging. Consequently, most research groups have abandoned the idea of pulse compression via filamentation.

Here we report on the generation of intense few-cycle pulses by efficient spectral broadening in a HCF by using intermediate intensities that are beyond the regime of pure Kerr-driven phase modulation, while remaining below the regime where a high degree of ionization occurs. In this regime, the mechanism of the spectral broadening can be influenced by adjusting the gas pressure. In this way we significantly improve upon the long-standing record of 5 mJ/5 fs [18] for kHz few-cycle pulse generation in the visible spectral range, and generate 3.8 fs pulses with 6.1 mJ pulse energy at 1 kHz repetition rate, implying a peak power exceeding 1 TW. In addition, the high-energy pulses are used for the generation of intense high-harmonic radiation in gases.

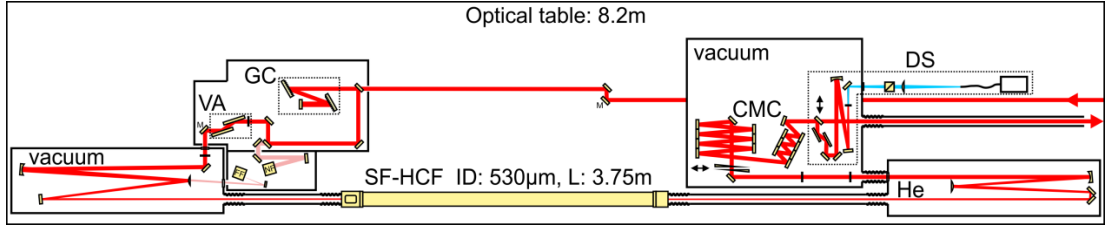

Fig. 1 The optical layout. GC: grating compressor, VA: variable attenuator, SF-HCF: stretched flexible hollow-core fiber, CMC: chirped mirror compressor, DS: dispersion scan device.

The experimental layout that we have realized at MBI (see Fig. 1) is an up-scaled version of a compressor that we have reported in [19]. After ~20 m of free propagation the uncompressed pulses of a KMLabs RedWyvern Ti:sapphire amplifier are compressed to 50 fs by a water-cooled grating compressor (GC). After passing a variable attenuator consisting of a half-waveplate and two subsequent thin-film polarizers the attenuated pulses are sent into a vacuum chamber, which contains a quarter-waveplate to convert the originally linearly polarized light to circular polarization followed by an astigmatism-compensated reflective focusing telescope with radii of curvatures of -3 m and 2 m, respectively. The leakage of the second telescope mirror is coupled out of the chamber for beam stabilization (TEM Aligna) resulting in an RMS fluctuation of the focal spot at the HCF entrance below 8 µm. We use a 3.75 m long SF-HCF as a waveguide with an inner diameter (ID) of 530 µm. Its length is solely limited by the 8.2 m length of the optical table and by the damage threshold of the mirrors around the HCF. The output side of the waveguide is connected to a gas-filled chamber (labeled "He" in Fig. 1) containing the re-collimating telescope mirrors with radii of 2.5 m and -3 m, respectively. At the output side of the HCF protected silver mirror coatings are used to support the broad spectrum of the pulses. The collimated beam is sent through a 3 mm AR-coated fused silica window to the compressor chamber, which is evacuated and which is the first element of the high-harmonic generation (HHG) vacuum beamline. Alternatively, a mirror can be inserted into the collimated beam in order to perform energy and beam profile monitoring outside the vacuum system. In the compressor chamber the polarization of the beam is first converted back to linear by an air-spaced achromatic quarter-waveplate before passing through a 2 mm thick z-cut KDP crystal used for optimizing the third-order dispersion (TOD) [20, 21]. A motorized AR-coated fused silica wedge-pair and a set of 14 chirped mirrors (PC70, Ultrafast Innovation) are used for fine adjustment of the group-delay dispersion (GDD). The compressed pulses are sent to the HHG setup or are deflected by a mirror mounted on a motorized stage towards a home-made second-harmonic generation (SHG) dispersion scan (d-scan) arrangement for in-vacuum pulse characterization.

In order to compress the pulses with maximal energy, systematic measurements were made to identify the best operation parameters. 50 fs pulses with up to 14 mJ pulse energy were used at the input of the HCF resulting in a peak power of up to 260 GW and a peak intensity of up to $4.4 \cdot 10^{14}$ W/cm², which is beyond the common operation range (up to $2.4 \cdot 10^{14}$ W/cm²) of a helium-filled HCF of 530 µm ID [6].

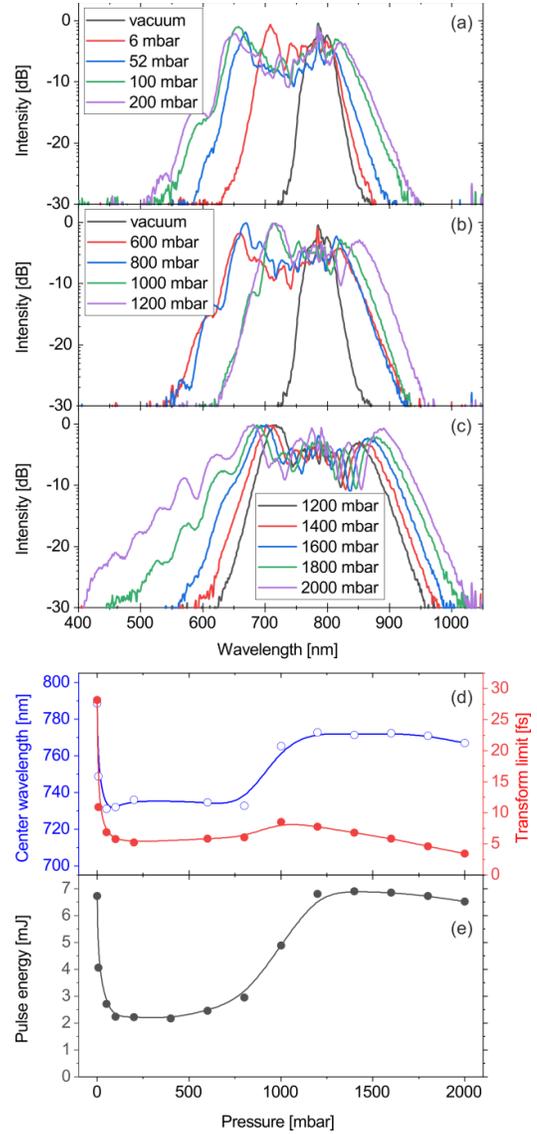

Fig. 2 Evolution of the spectral broadening measured at the output of the HCF with increasing gas pressure: **a.** regime of ionization-driven broadening at low pressure; **b.** transition between the ionization- and Kerr-induced broadening observed in an intermediate pressure range; **c.** Kerr-induced spectral broadening observed at higher pressures. **d.** center of gravity of the spectrum (blue line) and the transform limited pulse duration (red line) and **e.** the compressed pulse energy as a function of the gas pressure applied on the output side of the HCF.

A series of spectra recorded at increasing helium pressure applied on the output side of the HCF (see Fig. 2) reveals surprising behavior: up to ~200 mbar pressure the output spectra are very asymmetric and display a large blue shift which is attributed to ionization-induced broadening [22]. At the same time the beam profile becomes structured with a well-defined bright core and a large concentric ring around it and the energy after the compressor drops dramatically (see the left-hand side of Fig. 2.e). We note that the first mirror of the re-collimating telescope in the He-filled chamber has a 1" aperture which is large enough to reflect the beam emerging from an evacuated HCF. However, if the divergence of the beam increases then a considerable amount of energy is cut by the mirror. Consequently, the observed strong energy drop is not only due to ionization or ionization-induced waveguide losses, but also due to the limited aperture of the beamline behind the capillary.

At the intermediate pressure range (600-1200 mbar, shown in Fig. 2.b) the spectrum becomes gradually narrower and more symmetrical to the center wavelength of the input pulses. At the same time the output energy increases rapidly due to the energy redistribution inside the beam profile.

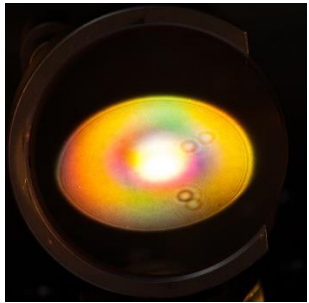

Fig. 3 True-color photograph of the collimated beam on the surface of a 2" silver mirror, when 2 bar of He was applied to the output side of the HCF.

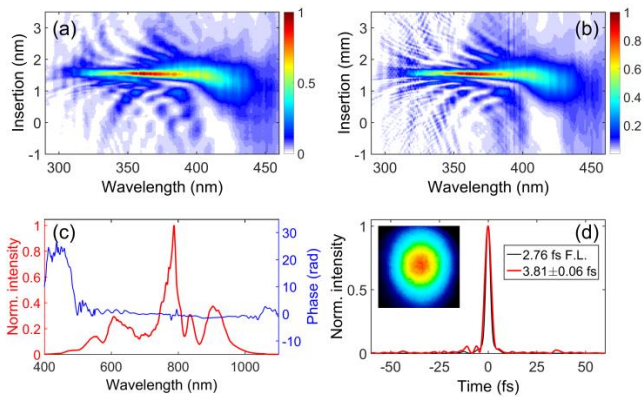

Fig. 4 D-scan measurement of the 6.1 mJ 1.5-cycle pulse. Panel **a.** and **b.** display the measured and retrieved traces, respectively. Panel **c.** displays the measured spectrum (red line) together with the retrieved spectral phase (blue line), while panel **d.** shows the retrieved temporal pulse shape (red line) together with the transform-limited shape (black line). The inset of panel **d.** shows the beam profile recorded by a linear detector.

At pressures above 1200 mbar (see Fig. 2.c) the spectral shape is characteristic for Kerr-induced broadening and the output energy increases back to the value obtained without gas in the HCF. The high transmission indicates that the ionization losses are balanced by a drop of the linear waveguide losses, due to better confinement of the light inside the waveguide. Further increasing the pressure above 1500 mbar even larger spectral broadening can be achieved. However, this goes at the expense of the transmitted energy. An attractive trade-off between bandwidth and energy is found at a 2000 mbar pressure where the spectrum is broad enough to support a transform-limited pulse duration of ~3 fs at more than 6 mJ pulse energy behind the chirped-mirror compressor. At this pressure, the beam profile still exhibits a very bright core and a homogeneous broad pedestal, as shown in Fig. 3.

After compression the pulses were characterized by an in-vacuum d-scan device (see Fig. 4). For the phase retrieval a differential evolutionary algorithm [23] was applied, resulting in very clean pulses with an FWHM duration of 3.81±0.06 fs. Here, the upper and lower bound of the pulse duration are given by the RMS deviation of 100 consecutive retrievals of the measured trace. The peak power was calculated by dividing the pulse energy by the integral of the retrieved pulse shape normalized to unity on the time window shown in Fig. 4.d resulting in 1.24 TW.

We used the compressed laser pulses for high harmonic generation (HHG) in a 10 cm long cell filled with argon or neon, using a loose focusing geometry (f = 5 m). First, when using argon, the energy of the generated harmonics was measured. The harmonics were first reflected off two consecutive $Nb_2O_5$ XUV-selective mirrors, in order to reject most of the fundamental radiation. After a grazing incidence reflection on a gold mirror the beam was further filtered by several Al filters (total thickness 200 nm) before impinging onto a calibrated XUV photodiode. We note that in the absence of the XUV-selective mirrors the Al filters were destroyed by the intense fundamental beam. Taking the transmission of the optics and the nominal quantum efficiency of the photodiode into account we estimate the pulse energy of the transmitted harmonics to be 260 nJ. We note that this value is very conservative, since it does not take the transmission drop of the Al filters due to oxidation and fast aging of the photodiode into account.

In order to generate harmonics up to 100 eV neon was used. Fig. 5.a displays the XUV spectrum recorded behind a 100 nm Al filter, whose cut-off at ~72 eV can be clearly seen. The spectrum consists of a series of broad harmonics which are overlapping beyond 60 eV forming a continuum. In order to observe the harmonic cut-off, the Al filter was replaced by a Zr filter. The transmitted spectrum shown in Fig. 5.b does not exhibit any modulations in the cut-off region in the case of optimally compressed driver pulses. When adding 0.14 mm of fused silica into the beam path of the driver pulses (an increase in the GDD and in the TOD of 5.5 $fs^2$ and 3.7 $fs^3$, respectively) the cut-off region becomes narrower and modulated (see Fig. 5.c), indicating the high sensitivity of the HHG process to the driver pulse shape.

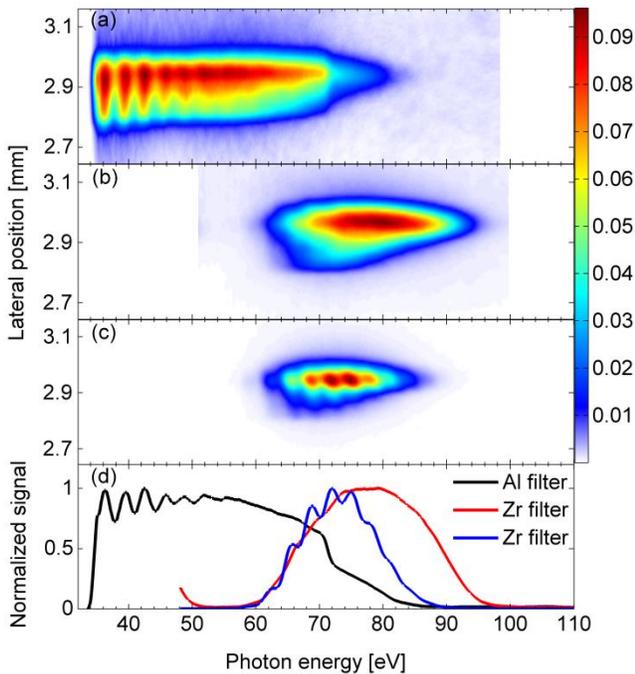

Fig. 5 High-harmonic spectra generated in a 10 cm long neon-filled cell, recorded behind an Al filter **a.** or behind a Zr filter **b.** and **c.** In **b.** the driver pulses are optimally compressed, while in **c.** the driver was slightly positively chirped. Panel **d.** shows the spatially integrated spectra for all three cases.

In conclusion, we have demonstrated HCF compression in an intermediate intensity regime between ionization-induced and Kerr-dominated spectral broadening and have observed the transition between these two regimes by changing the gas pressure. We have demonstrated the generation of 6.1 mJ 3.8 fs pulses centered at 764 nm which represents the highest energy sub-2-cycle pulse ever achieved in the visible range at kHz repetition rate. Compared to our recent work [19] starting with two-times longer pulses of slightly higher energy we achieve almost the same pulse duration and nearly a factor of two higher compressed energy.

We have demonstrated the generation of intense high-harmonics with the pulses, showing continuous spectra in the cut-off region in both argon and neon gases, which is a strong indication that an isolated attosecond pulse is generated. In the future, two-pulse autocorrelation measurements will be implemented, in order to characterize the pulse duration of the harmonic pulses. This provides a route towards attosecond nonlinear XUV spectroscopy.

**Acknowledgment**. We thank for valuable discussions with A. Tajalli (Leibniz Universität Hannover) concerning the d-scan measurements and with T. Witting regarding the harmonic generation. We are indebted to G. Steinmeyer and P. Simon (Laser-Laboratorium Göttingen) for their suggestions concerning the manuscript.

**Disclosures**. TN: Laser-Laboratorium Göttingen e.V. (P).